\documentstyle[aps,amssymb,epsfig,pre]{revtex}

\begin{document}
\draft

\title{Swelling of phospholipid floating bilayers: the effect of chain length}

\author{Giovanna Fragneto$^1$\thanks{to whom correspondence
should be addressed at fragneto@ill.fr}, Thierry
Charitat$^2$, Edith Bellet-Amalric$^3$, Robert Cubitt$^1$, Fran\c cois
Graner$^4$
}

\address{$^1$ Institut Laue-Langevin, 6 rue Jules Horowitz, B.P. 156, 38042
Grenoble Cedex, France\\
$^2$ Institut Charles Sadron \thanks{ CNRS - UPR 22, Universit\'e Louis
Pasteur}, 6 rue Boussingault,
F-67083
Strasbourg
Cedex, France\\
$^3$ DRFMC/SP2M/SGX, CEA,17 Av. des Martyrs 38054 Grenoble Cedex 9,
France\\
$^4$ Spectrom\'etrie Physique \thanks{ CNRS - UMR 5588,
Universit\'e Grenoble I}, B.P. 87,
F-38402 St Martin d'H\`eres Cedex, France}

\date{\today}
\maketitle
\begin{abstract}

The equilibrium distance between two lipid bilayers stable in bulk water and in
proximity of a substrate  was investigated. Samples consisted of a
homogeneous lipid bilayer, floating near an identical bilayer
deposited on the hydrophilic surface of a silicon single crystal. Lipids
were saturated
di-acyl phosphocholines, with the number of carbon atoms per chain, n,
varying from 16 to 20. The average and r.m.s. positions of the floating
bilayer were determined by means of neutron specular reflectivity. Samples
were prepared at room temperature (i.e. with the lipids in the gel phase)
and measurements performed at various temperatures so that the whole region of
transition from gel to fluid phase was explored. Data have been interpreted
in terms  of competition between the interbilayer potential and membrane
fluctuations and used to estimate the bending rigidity of the bilayer.

\end{abstract}




\bigskip

\section{Introduction}

The basic building blocks of cell membranes are lipid bilayers. They provide
mechanical stability and have a strong tendency to form closed structures.
They support membrane proteins and have sufficient flexibility
for vesicle budding and membrane fusion \cite{gennis}. The physical
properties of lipids are fundamental for understanding their biological
functions. Because of their complexity, the duplication of cell membranes
and the investigation of their interactions with peptides or proteins is
difficult. An increasing need for biomimetic models has recently
caused the revival of a classical approach: the use of lipid bilayers
\cite{katsaras}. A considerable effort has led to numerous successes in
preparing new types of samples and corresponding structural measurement
techniques. These can resolve sub-nanometric details, leading to
investigations of lipid-lipid, lipid-peptide or lipid-protein interaction
mechanisms.

Phospholipids present a variety of different phases as a function of
temperature \cite{liporevue}. Cooling from high to low temperature, lipids
overcome the fluid to gel phase transition at a temperature commonly known
as T$_m$ (melting temperature). While in the fluid phase, $L_\alpha$, lipid
chains are in a liquid-like conformation and mobile in the lateral
direction, in the gel phase, $L_\beta$, chains are stiffer. For some
lipids, just below T$_m$, there is the so-called ripple phase,  $P_\beta$,
where the lamellae are deformed by a periodic and static out of plane
modulation.
Biophysical studies of membrane-membrane and
membrane-protein interactions
require well controlled model systems \cite{katsaras,mouritsen}. In
previous papers \cite{charitat,fragneto} a new model system has been
described consisting of stable and reproducible double bilayers, in which a
lipid bilayer floats at 2-3 nm on top of an adsorbed one. Since the second
bilayer is only weakly bound, the preparation is delicate but results are
reproducible. The process relies on three vertical Langmuir-Blodgett
depositions, followed by a horizontal Langmuir-Schaeffer one; this method is
efficient when lipids are in the gel phase and by raising the temperature
the floating bilayer overcomes a phase transition becoming fluid and still
stable.

The floating bilayer system has been very useful to probe bilayer-bilayer
interactions, and is currently being improved to accommodate for charges in
water and in the bilayer, so that lipid/peptide interactions can be studied
in physiological buffers. Since it has the potential to support
transmembrane proteins, this system can be a better biomimetic model
membrane than supported single bilayers.
Floating bilayers offer the
following advantages: they are well-defined
structures, where the composition of each leaflet of the bilayer can be
chosen separately; they are immersed in aqueous solution and they allow
information on a single bilayer to be obtained  \cite{heitz,mouritsen2},
contrarily
to the classical models of multilamellar vesicles or stacked systems. They
are stable in the fluid phase and swelling of the
water layer between the bilayers can be observed with this system around
the phase transition.

A systematic study of the swelling is presented here, where the values of
the distance of the floating bilayer from the adsorbed one were determined
for
phosphocholine lipids with saturated double chains and a number of carbon
atoms, n, per chain from 16 to 20. For all samples a more or less broad
region around the main phase transition temperature has been detected
characterised by a swelling of the layer of water between the bilayers of
about 1 nm. Giant swelling (i.e. swelling $\gg$ 1nm) was also observed with
samples having 17 and 18 carbon chains. Results from a self-consistent
theory \cite{mecke}, that has been developed to extract the values of the
bending modulus of the membranes from the experimental data, will be shown.

\section{Materials and Methods}

Lipids were purchased from Avanti Polar Lipids (Lancaster, Alabama, USA)
and used without further purification. They are phosphatidyl-cholines (PC)
with different chain lengths (for simplicity they are indicated as
Di-C$_{n}$-PC): di-palmitoyl (Di-C$_{16}$-PC), di-heptadecanoyl
(Di-C$_{17}$-PC), di-stearoyl (Di-C$_{18}$-PC), di-nonadecanoyl
(Di-C$_{19}$-PC) and di-arachidoyl (Di-C$_{20}$-PC). Fully deuterated
(Di-C$_{18}$-PC) (d83) was also used. Several attempts to prepare floating
bilayers from fully deuterated Di-C$_{16}$-PC failed. Hydrogenated lipids
were measured in D$_2$O and fully deuterated lipids in H$_2$O and
Silicon Match Water, (i.e. mixture of H$_2$O and D$_2$O with a
scattering length density = 2.07x10$^{-6}$\AA$^{-2}$). The minimum and
maximum temperatures investigated are: 25.0-82.6$^{\circ}$C for n=16;
25.0-55.6$^{\circ}$C for n=17; 25.0-64.1C for n=18; 25.0-60.8$^{\circ}$C
for n=19 and 25.0-68.6$^{\circ}$C for n=20.

Two kinds of silicon substrates were used: (i) of surface 5x5 cm$^2$ and 1
cm thick obtained and polished by SESO (Aix-en Provence, F) with a r.m.s.
roughness 0.15 nm measured on 300x300 $\mu$m$^2$; and (ii) of surface
8x5 cm$^2$ and 2 cm thick obtained from Siltronix
(Archamps, F) and polished by the Optics Laboratory of the European
Synchrotron Radiation Facility (Grenoble, F) with a technique including
diamond powder treatment and producing high quality crystal surfaces with
r.m.s. roughness of $<$0.2 nm.

The silicon blocks were made highly hydrophilic with a UV/ozone treatment
as in \cite{fragneto}. Lipids were deposited on the substrate with a
combination of the Langmuir-Blodgett and Langmuir-Schaeffer techniques as
described in \cite{charitat,fragneto}. Depositions were done at room
temperature, and the lipids compressed to surface pressures varying with
the lipid chain length (40 mN/m for n=16,17,18; 35 mN/m for n=19 and 30
mN/m for n=20). Transfer ratios for the first layers were always high and
of the same order as samples described in \cite{fragneto}. They improved as
the chain length increased and attained the value of 1.0 for n=18,19,20. It was
found that the horizontality of the sample in the Schaeffer dip is
crucial for a good quality sample. In fact, while in
\cite{charitat,fragneto} during the Schaeffer dip the substrates were
slightly inclined so that no air bubbles would form on the solid surface,
here systematic measurements on Di-C$_{18}$-PC samples were made and it was
found that a sample surface as parallel as possible to the water surface
(micrometric
screws were used) prevents the formation of bubbles and gives transfer
ratios always bigger than 0.97.

While it is difficult to prepare
samples with short phospholipid chains (e.g. with DMPC), it has been found
that this is possible if the first layer is chemically grafted to the
surface \cite{hughes}, although in this case the coverage of the DMPC
floating layer is not always optimal and reproducibility problems are
encountered. This method when applied to longer chain lipids has resulted
in the formation of very stable and defect-free layers (work in progress).

\section{Neutron reflectivity: basic principles and measurements}

Specular reflectivity, R(Q), defined as the ratio between the specularly
reflected and incoming intensities of a neutron beam, is measured  as a
function of the wave vector transfer, $Q = 4\pi \sin{\theta}/\lambda$
perpendicular to the reflecting surface, where $\theta$ is the angle and
$\lambda$ the wavelength of the incoming beam. R(Q) is related to the
scattering length  density across the interface, $\rho(z$), and for values
of Q bigger than the critical value for total reflection, it decays as
$Q^{-4}$ \cite{penfold}.

Specular reflectivity allows the determination of the structure of matter
perpendicular to a surface or an interface. Experiments are performed in
reflection at grazing incidence. Samples must be planar, very flat and with
roughness as small as possible ($<$0.5 nm).
The data presented here were collected at the High Flux Reactor (HFR) of
the Institut-Laue
Langevin (ILL, Grenoble, F) both on the small angle diffractometer D16 and
the high flux reflectometer D17.
Instrument set-up and the data collection method on D16 have been described in
\cite{charitat}. A monochromatic neutron beam of wavelength $\lambda$= 4.52
\AA \ impinged on the sample at grazing incidence. A two-dimensional $^3$He
wire detector was kept in a fixed position covering the angular 2$\theta$
range 0$^{\circ}$- 9$^{\circ}$. The beam was vertically focused at the
detector position in order to gain intensity without worsening divergence.
The dual mode instrument D17 \cite{cubitt1,cubitt2} is designed to take
advantage of both Time-Of-Flight (TOF) and monochromatic methods of
measuring reflectivity. In this study measurements were all taken with the
instrument in the TOF mode by using a spread of wavelengths between 2 and
20\AA \ at two incoming angles (typically 0.7$^{\circ}$ and 4$^{\circ}$)
and with a time resolution, defined by two rotating choppers, of $\Delta t /t
=1\%$ and 2-5$\%$, respectively, where $\Delta t$  is the neutron pulse
duration. The beam was defined in the horizontal direction by a set of two
slits, one just before the sample and one before a vertically focusing
guide.
Neutron reflectivity from the same sample was measured at different
temperatures monitored with a thermocouple (equilibration time 25
minutes, stability $<0.1^{\circ}$C, absolute precision
$<0.3^{\circ}$C), in the water-regulated sample
chamber already described in \cite{charitat}. Two types of measurements
were performed: (i) up to Q $\sim$ 0.08
\AA$^{-1}$; (ii) over the whole available Q-range defined as follows.

For measurements done on D16, the whole available Q-range
(0.003-0.22 \AA$^{-1}$) was scanned in eight hours, after equilibration. On
D17 the useful Q-range, before hitting the sample background, spanned from
0.007 to 0.25 \AA$^{-1}$ and was measured in about two hours, also after
equilibration. The equilibration time varied depending on the step of
temperature increase or decrease. For a step of 1$^{\circ}$C it was of
about 15 minutes.

\section{Data analysis: model fitting of reflectivity data}

A reflectivity curve essentially reflects the sample density perpendicular
to the substrate surface, or rather the square modulus of its Fourier
transform. Data analysis requires a good knowledge of the sample and
here it was done via model fitting. Born and Wolf give a general solution,
the so called optical matrix method \cite{born}, to calculate the
reflectivity from any number of parallel, homogeneous layers, which is
particularly useful since any layer structure can be approximately
described by dividing it into an arbitrary number of layers parallel to the
interface, each having a uniform scattering length density.
In previous studies \cite{fragneto,penetratin}, multiple contrast neutron
measurements at room
temperature had allowed the determination within \AA \ precision of the
profile of adsorbed
and floating bilayers. Each bilayer is resolved into outer-inner-outer
slabs: outer slab = heads (phospho-choline and glycerol groups), inner slab
= tails (hydrocarbon chains). A water film of thickness D$_w$  separates
the adsorbed and floating bilayers. Together with the natural oxide layer
present on the silicon crystal and the hydration water layer between the
substrate and the first lipid layer, this leads to 9 slabs in total, each
one having its own average thickness and scattering length density.
Each interface between two slabs has a certain width, also called
r.m.s. roughness. Since this is included into the density profile, it can be
determined by fits of reflectivity curves, but it is not possible to
determine whether it is a static intrinsic width or an average over the
neutron beam correlation length of temporal or spatial fluctuations.

The present work focuses on thicknesses and roughnesses only. By starting
from the above 9-slab model, with the tabulated values of $\rho$ (see
Tables in refs. \cite{charitat,fragneto}) the relevant parameters to
the data were fitted for all systems in the gel phase (the starting point
of all
measurements was the collection of the reflectivity profile at
25$^{\circ}$C). The first reflectivity minimum abscissa is very
sensitive to the thickness of both water films. The second reflectivity
minimum position is mainly sensitive to the thickness of the first bilayer
tail slab; the amplitudes and sharpnesses of both minima depend upon bilayer
roughness (see figures 1 and 2). All slabs were easily distinguished,
except that the high hydration of lipid heads \cite{jendrasiak,naglehydr}
prevents a localization of the heads/D$_2$O interface to better than 0.1
nm. Correcting the scattering length densities for temperature variations
had no significant effect on the positions of the reflectivity minima,
hence on the fitted parameters. For data at high temperatures the analysis
concentrated
on the exact determination of D$_w$. When giant swelling occurred, data
analysis was complicated by the fact that the high values found for
the roughness limit the usefulness of the analysis based on homogeneous
layer models, leading to large error bars for D$_w$ and $\sigma$.

In a parallel study (in progress) of specular and off-specular reflectivity
from synchrotron radiation on similar samples  the wider
q-range attained and
therefore the inadequacy of the box model fitting \cite{Shalke} has made it
necessary to use the one Gaussian hybrid model described in
\cite{nagle1996}. The results as far as D$_w$ is concerned are in perfect
agreement with the values obtained from neutron reflectivity using the box
model.

Figure 1 gives some examples of the collected data. The reflectivity curves
obtained from hydrogenated lipids in the fluid phase at randomly chosen
temperatures, as well as the best fits to the data, are shown. As expected
the main differences are observed at the second minimum which is dependent
on the single bilayer structure. Figure 2 is an example of curves collected
for the system Di-C$_{17}$-PC in the gel and fluid phases and in the
transition region, as well as the best fits to the data. The density
profiles obtained from the model that best fits the data are given in the
insert.

\section{Data analysis: calculation of bending modulus}

\label{theory}

The self-consistent theory described in \cite{mecke} for a
fluid bilayer floating near a solid substrate was applied to the results of
this study. The basic principles are briefly described below. The shape of
the substrate potential determines the amplitude $\sigma$ of thermal
fluctuations which itself determines the effective potential
\cite{liporevue,helfrich} felt by the bilayer. The average bilayer
position and its r.m.s.
fluctuation amplitude are thus coupled and depend both on temperature and
bending energy of the membrane. This model situation is experimentally
relevant as already reported in \cite{mouritsen}. Three self-consistent
equations may be written for the partition function, ${\cal Z}$,
the equilibrium inter-bilayer distance, D$_w$, and the r.m.s. fluctuation
amplitude, $\sigma$:

\begin{eqnarray}
\label{self}
\nonumber
\displaystyle {\cal Z} &\ = \ &\displaystyle \int dz \; e^{-16
\frac{\kappa}{(k_B T)^2} \sigma^2
U(z) - 3 \frac{(z - D_w)^2}{8 \sigma^2}}, \\
\displaystyle D_w & \ = \ &\displaystyle  \frac{1}{{\cal Z}} \int dz \;
z\;  e^{-16\frac{\kappa}{(k_B
T)^2}\sigma^2 U(z) - 3\frac{(z - D_w)^2}{8\sigma^2}}, \\
\nonumber
\displaystyle \sigma^2 & \ = \ &\displaystyle  \frac{1}{{\cal Z}} \int
dz \; (z - D_w)^2 \;
e^{- 16 \frac{\kappa}{(k_B T)^2} \sigma^2 U(z) -
3\frac{(z - D_w)^2}{8\sigma^2} }\;.
\end{eqnarray}

where $k_B$ is the Boltzmann constant, T the temperature and $\kappa$ the
bilayer bending rigidity modulus. U(z) is a one dimensional potential
describing the interaction between the bilayer and the substrate.

At short distances, with $z$ typically of the order of the hydration length
$z_0
\sim$ 0.6 nm or
smaller, the potential is dominated by the hydration repulsion due to water
molecules inserted between hydrophilic lipid heads \cite{liporevue}~:

\begin{equation}
\label{hydratation}
U_h(z) = A_h \; \exp \left( -2\frac{z}{z_0}\right).
\end{equation}

At long distances, for a neutral membrane,  the dominant term in U(z) is due
to van der Waals attraction~:
\begin{equation}
\label{vdw}
U_{vdW}(z) = - \frac{A}{12 \pi} \left[ \frac{1}{z^2} - \frac{2}{(z +
\delta)^2
} +
\frac{1}{(z + 2 \delta)^2} \right],
\end{equation}
where $\delta$ is the bilayer thickness and $A$ the Hamaker
constant.

Here, it is assumed that $A$ does not diverge and does not undergo
any marked minimum so that its variations with temperature are negligible
(compared to those of $\kappa$) and therefore $A$ is constant in what follows.

The dimensionless parameter $\beta$~:
\begin{equation}
\label{beta}
\beta = \frac{(k_B T)^2}{A \kappa} \;  ,
\end{equation}

determines the position and the fluctuations of the membrane. At low values
of $\beta$, the bilayer is stiff, barely fluctuating
in a strong minimum of potential energy. At high values of $\beta$, the
bilayer is soft and has large fluctuations leading to the increase of the
mean distance D$_w$ to the adsorbed bilayer.

By eliminating ${\cal Z}$ from the three self-consistent equations
(\ref{self}), two of the three physical quantities  D,
$\sigma$ and $\beta$ can be expressed as a function of the third one.
Experimentally,
D$_w$ and $\sigma$ are measurable quantities while $\beta$ is
much more difficult to measure, especially for bilayers in the gel phase
\cite{koenig}. Therefore, measurements of D$_w$ are used here to estimate
$\beta$ and $\sigma$ by
using a classical numerical resolution of equations (\ref{self})
\cite{mecke}. The validity of this approach is tested by comparing
$\sigma$ to the experimental data.

\section{Results and discussion}

The analysis of the reflectivity curves has led to the determination of the
water layer D$_w$ and the bilayer roughness $\sigma$ (see section IV). The
main results are
summarised in Table I as well as in Figures 3 to 9.
It was found that the silicon
crystals were covered by a 1.2$\pm$0.1 nm thick
oxide. A thin water film of thickness 0.6$\pm$0.1 nm separated it
from the adsorbed bilayer. Both in the gel and in the fluid phases the
thickness of the headgroups was 0.8$\pm$0.1 nm. The thickness of chains
(D$_{chain}$) in the gel phase (at 25$^{\circ}$C) is reported in Table
I for all lipid species. In the fluid phase those thicknesses
were usually lower by
about 0.2 nm. The bilayers were found to incorporate 5 to 10$\%$ of water.
As for the roughness, well below and above the transition region it was the
same as the substrate, i.e. 0.3$\pm$0.1 nm. In the transition region it
increased (see below).

Figures 3 and 4 are a summary of the changes of D$_w$ with temperature for
all lipid species. Figure 3 refers to measurements from hydrogenated lipids
in D$_2$O and Figure 4 refers to measurements of deuterated Di-C$_{18}$-PC
in H$_2$O and contrast match water. On the x-axis is T-T$_m$, i.e the
difference between the temperature at which the measurement was taken and
the value of the melting temperature (the values used for T$_m$ are listed
in Table I). For all lipids a very similar behaviour was observed, that is
an increase of both D$_w$ and $\sigma$ in a certain range of temperature
around the main phase transition (swelling region). The degree of swelling
was always around 1 nm, and in two cases (n=17,18) giant swelling ($\gg$1
nm) was observed. The
swelling temperature T$_s$ defined as the temperature at which D$_w$ and
$\sigma$ are maximun appears to be systematically 4 to 5 degrees below the
melting temperature (see also figure 9).  Above the transition, the
floating bilayer is stable and D$_w$ goes back to values similar to those
found in the gel phase, although slightly higher. This equilibrium distance
in the fluid phase increases slowly with temperature. In the gel phase, the
minimum value D$_{w,g}$, listed in Table I, is obtained at the lowest
measured temperature (i.e. 25$^{\circ}$C). This value seems to be constant
over a large temperature range, indicating that it is probably a good
estimation of the bilayer position in the absence of thermal fluctuations.
Figure 5 shows the variation of this minimum value with the lipid
chain length n. At 25$^{\circ}$C, D$_{w,g}$ tends to decrease when n
increases, and the effect seems to depend also on the oddness or evenness
of n. It is unclear if this is fortuitous, to the best of the authors
knowledge there
are no other examples in the literature.

\begin{table}[ht]
\begin{center}
\label{table1}
\caption{List of experimental data and literature parameters for the lipid
species studied. n is the number of carbon atoms per chain in the formula
Di-C$_n$-PC; T$_p$ is the temperature of transition from the gel to the
ripple phase \protect\cite{lipidat}, T$_m$ the temperature of main
transition \protect\cite{lipidat}.
In column 4 are listed the regions of temperature where swelling was observed,
T$_s$ is temperature at which D$_w$ reaches  its  maximum value D$_w$ (max)
and D$_{w,g}$ is the value of D$_w$ in Gel phase. D$_{chain}$ is the
chain region thickness, for the
floating bilayer, in the gel phase (25$^\circ$C). $A_h$ and $A$ are defined
in the text and estimated from D$_{w,g}$ (see below).}
\begin{tabular}{|c|c|c|c|c|c|c|c|c|}
n &    {T$_p$}     &   {T$_m$}   & Swelling Region   &  T$_s$ (max) &
D$_{chain}$ & D$_w$ (max) & D$_{w,g}$ & $\frac{A_h}{A}$\\
    & ($^\circ$C)             &       ($^\circ$C)    &     ($^\circ$C)
&    ($^\circ$C)  &  (nm) &  (nm)                 & (nm)   &   \\
\hline
16 & 35-37             & 41-41.2 & (35-40)$\pm$1 &  & 3.2$\pm$0.2  &
3.3$\pm$0.2  & 2.2 $\pm$ 0.1 & 80 $\pm$ 15\\
      &  &  (41)$^2$                     &                             &
&          &               &                   &       \\
\hline
17  & 43 & 48-49.8 & (40-45)$\pm$1 & 41.5$\pm$0.5 & 3.4$\pm$0.2  &
4.8$\pm$0.2 & 1.4 $\pm$ 0.1 & 23 $\pm$ 3\\
                         &     &         (48)$^2$    &
&   &                      &  &                      &  \\
\hline
18  & 50-52 & 54.6-55  & (45-55)$\pm$1  & 51.5$\pm$0.5 & 3.6$\pm$0.2  &
6.0$\pm$0.4  & 1.8 $\pm$ 0.1& 41 $\pm$ 7 \\
                       &       &  (55)$^2$
&     &    &  &     &  & \\
\hline
18 (D)\footnote{Fully deuterated lipid;$^2$ Value given by Avanti polar
Lipids and used for the figures;}  & &  & (37-50)$\pm$1  & & 3.6$\pm$0.2  &
2.8$\pm$0.1 & 1.8 $\pm$ 0.1 & 41 $\pm$ 7\\
                              &              &  (50.5)           &  &     &
&    &   & \\
\hline
19 & 55.7-57.8 & 59-60 &  (50-62)$\pm$1 & & 3.8$\pm$0.2 & 2.3$\pm$0.1 &1.2
$\pm$ 0.1 & 19 $\pm$ 2\\
                              &    &               (60)$^2$         &     &
&  & & & \\
\hline
20  & 62.1-63.7 & 63-66.4 &  (56-64)$\pm$1  &  & 4.0$\pm$0.2  & 2.5$\pm$0.1
&1.5 $\pm$ 0.1 & 26 $\pm$ 3 \\
                              &  &                (66)$^2$          &     &
&  &  & & \\
\end{tabular}
\end{center}
\end{table}

The reproducibility of the observed effects was checked on different
samples, from different experiments carried out in a period of two years on
various silicon substrates and instruments (either D16 or D17). Figures
6(a) and 6(b) show results from samples with n=16 and 18. For clarity
results have been split in different graphs so that the most notable
effects could be highlighted. For the n=18 case data from 4 samples are
reported (samples A, B, C, and D in the caption). The temperature range
where the swelling was observed is reproducible. The amplitude of swelling,
that is the maximal distance reached, is variable from one sample to the
other. In two cases giant swelling was observed, i.e. with lipids having
n=17 and n=18, the last one already described in \cite{fragneto}. It is not
excluded that giant swelling occurs also with the other lipids investigated
in this geometry. The limited available beam-time has not allowed exploring
the transition temperature ranges in fine detail for all samples and since
giant swelling is usually observed in narrow T ranges \cite{heimburg,pabst}
it might have been missed in other cases.\\

The reversibility of the swelling was also studied. Figures 7(a) and 7(b)
show results from one sample with n=17 and two samples with n=18 with data
collected while heating up and cooling down once or twice, as indicated in
the figure caption. The major result is that the swelling is reversible. On
both figures, a weak increase in the maximum amplitude of the swelling with
the successive heating and cooling processes is observed. It could still be
due to the narrowness of the temperature range where swelling occurs, as
mentioned above or it could be an indication that some defects of the
bilayer like pinning points have been annealed. The swelling
temperature
is barely sensitive to cycles in temperature~: T$_s$
appears to be slightly lower
when cooling down than when heating up. This could be related to the
hysteresis of the main transition although the effect is too weak for a
definitive interpretation.

Ref. \cite{mecke} suggests a possible interpretation of the swelling in
term of a balance between attraction by the substrate and entropic
repulsion due to thermal fluctuation.  The self-consistent theory briefly
described in section V enables expressing D$_w$ and $\sigma$ as function of
$\beta$ by assuming that the dimensionless
parameter $\frac{A_h}{A}$ and the characteristic length $z_0$ of the
microscopic potential (equations \ref{hydratation}
and \ref{vdw}) are known. Here it was assumed that $z_0$ is equal to 0.6
$nm$ and that it does not depend on chain length.
By using the experimental value
of D$_w$ at low temperature (D$_{w,g}$ in table I) and the
microscopic potential, the ratio $\frac{A_h}{A}$ was estimated for
different values of n (Table I). From this approach, the
experimental values of D$_w$
are used to estimate $\beta$ and the corresponding value of $\sigma$ for
each temperature. The value of $\sigma$ is then compared to the r.m.s.
roughness obtained by model fitting the data. Results from one
representative example (i.e. n=16) are given in figure 8. Agreement with
all samples is remarkable (data not shown).

The derivation of an experimental estimation of $\kappa$ implies the
knowledge of the energy scale A (Hamaker constant). This value was
estimated for all samples by normalising $\kappa$ to the usual value 10$k_B
T$ in the fluid phase. Values of A of the order of
$10^{-21}-10^{-22} J$ were found in agreement with
literature data \cite{liporevue}. This method forces the values for
all different
lipids to overlap artificially at high temperature.
Figure 9 shows the value of the bending modulus calculated in that way. A
minimum of the bending modulus is found around the main phase transition.
Such minimum in $\kappa$ is in agreement with literature data
\cite{meleard,mishima,pouligny,lee}, even if its physical origin
is unclear. It could be
a signature of the proximity to a critical
point in the phase diagram. As already mentioned, for
Di-C$_{18}$-PC in D$_2$O the pre-transition temperature T$_p$ =
51-52$^{\circ}$C \cite{guard,ohki} is close to the temperature at which the
largest swelling is observed here. On the other hand, a minimum in $\kappa$
is more likely to exist near the main (melting) transition T$_m$
\cite{pouligny,lemmich1}. This would imply that for the floating bilayer
T$_m$  is lower than the value reported for multilamellar systems
\cite{guard,ohki}. A comparable lowering of the melting temperature has
been previously observed on single bilayers on a solid support \cite{bayerl}
and was attributed to surface tension effects. Two experimental
facts enforce this hypothesis:
(i) the swelling temperature T$_s$
seems to be always shifted
by 4-5$^{\circ}$C from the bulk T$_m$ value (figure 3 and 9) independently
of the phospholipid chain length; (ii) measurement made on deuterated
Di-C$_{18}$-PC in H$_2$O in water exhibit also swelling and no literature
data were found on a transition to the ripple phase for those deuterated
species.
In floating bilayers, a lateral stress
might arise from possible pinning on defects. However, such a large tension
would strongly limit the fluctuations and prevent both D$_w$ and $\sigma$
from increasing as much as we observe \cite{lipoleible1,perino}.
Finally, this interpretation can explain why the equilibrium distance is
highly sensitive to small variations in the bending modulus and in the
balance between attraction and repulsion (the samples are on the verge of
unbinding) \cite{lipoleible1,lipoleible2}.

\section{Conclusions and perspectives}

Floating bilayers offer interesting advantages over other models for
membranes. A structured bilayer, stable in time and free to fluctuate in
the fluid phase enables in-situ studies of transmembrane channels or
interactions between a lipid bilayer and membrane proteins.  Its size,
orientation, and the proximity of a smooth substrate enable reflectivity
studies, and therefore structural characterisation in the fraction of
nanometre sca le. The composition of both lipids and surrounding medium can
be varied with a certain freedom. Furthermore, this is a well defined
geometry for comparison with theoretical models; it also provides different
control parameters (temperature, composition of lipids or water) to study
the interaction between two membranes. Thanks to the excess water, a large
increase in both the floating bilayer roughness and the inter-bilayer
distance has been observed around T$_m$ for a series of samples from
phosphocholine lipids having different chain lengths. Anomalous swelling
has been observed frequently at the main phase transition in multilamellar
systems and it is object of debate \cite{lemmich2,nagle1} whether it is due
to an increase of the bilayer thickness or to the swelling of the water
layer between the membranes. The results presented here can help in this
debate since the swelling that we observe is undoubtedly due to the increase
in thickness of the water layer between the two bilayers.
A parallel study is in progress for the determination of the fluctuation
spectrum of the floating bilayers by using synchrotron radiation
off-specular reflectivity measurements. In fact, while specular reflection
gives information in the direction perpendicular to the interface, the
lateral structure of the interface may be probed by the nonspecular
scattering measured at reflection angles different from the specular one.

Since in that case the intensity of the reflected beam is much weaker than
in the specular direction, the use of synchrotron radiation is necessary
for obtaining a wide enough dynamic range for the analysis of these thin
systems.
Measurements have been performed both on double bilayers and on samples
prepared
by depositing three lipid layers on hydrophobically grafted silicon
substrates. In this way the fluctuations of the layer closer to the
substrate are reduced and information can be obtained only from the
floating one. Results (to be published) indicate that the layer is under
some tension (of the order of 3-5 mN/m) and such tension could explain why
we observe the swelling at values of temperature lower than literature
values of T$_m$.

Besides the bilayer fluctuation studies, the model is being improved to
make it more relevant for interactions with biological systems. So far, we
have succeeded inserting various amounts of cholesterol ranging from 0.5 to
10$\%$ cholesterol in the bilayers as well as negative charges (work in
progress).  The preparation of asymmetric bilayers, with the internal
leaflet formed by phospho-ethanolamine headgroups and the outer one by
phosphatidyl-choline headgroups is being optimised.\\
  \\
{\bf Acknowledgments :}

The authors wish to thanks A. Freund and the staff of the Optics
Laboratory at the ESRF for polishing the silicon crystals. We
thank
J. Daillant, A. Braslau (who was also present at one of the experiments),
K. Mecke for useful discussions, J. Katsaras for suggesting the
measurements at very high temperature, B.Stidder
for help in sample preparation and J. Allibon for optimising the
temperature control software that allowed for some sleep during the nights.

\newpage

\section{Figure captions}

\begin{figure}
\label{reflectivite1}
\caption{Reflectivity data (points) and best fits to the data (continuous
lines) from all lipid species, Di-C$_n$-PC, in the hydrogenated form and
measured in D$_2$O, in  the fluid phase: ($\bullet$) n=16  at
T=44.2$^{\circ}$C, ($\circ$) n=17 at T=45.8$^{\circ}$C, ($\blacksquare$)
n=18  at T=55.4$^{\circ}$C, ($\square$) n=19  at T=58.9$^{\circ}$C,
($\lozenge$) n=20  at T=68.6$^{\circ}$C.}
\end{figure}

\begin{figure}
\label{reflectivite2}
\caption{Reflectivity profiles (points) and best fits to the data
(continuous lines) profiles for a double bilayer of Di-C$_{17}$-PC at
($\bullet$)  T=25.0$^{\circ}$C, ($\circ$)  T=43.0$^{\circ}$C  and
($\square$)  T=45.8$^{\circ}$C. In the insert are the scattering length
density profiles as from the model used to fit the data: continuous thin
line for T=25.0$^{\circ}$C, continuous thick line for T=43.0$^{\circ}$C and
dashed line for T=45.8$^{\circ}$C}
\end{figure}

\begin{figure}
\label{resumeDw}
\caption{Distance, D$_w$, between the two bilayers from measurements from
hydrogenated lipids vs. the deviation of temperature from the melting
temperature, T-T$_m$;
($\bullet$) n=16; ($\circ$) n=17; ($\blacksquare$) n=18; ($\square$) n=19;
($\lozenge$) n=20.
Lines between points are added for clarity. Samples with n=16 and 18 were
measured on D16; samples with n=17,19 and 20 were measured on D17.}
\end{figure}

\begin{figure}
\label{DwHyDeut}
\caption{Distance, D$_w$, between the two bilayers from measurements from
deuteriated Di-C$_{18}$-PC in Silicon Match Water ($\circ$) and H$_2$O
($\bigtriangleup$) and for hydrogenated Di-C$_{18}$-PC in D$_2$O
($\blacksquare$) (same datas as figure 3)  vs. the deviation of temperature from the melting
temperature, T-T$_m$.}
\end{figure}

\begin{figure}
\label{parite}
\caption{Variation of D$_w$ as a function of n at T=25$^{\circ}$C.}
\end{figure}

\begin{figure}
\label{repro}
\caption{(a) Distance, D$_w$, between two Di-C$_{16}$-PC bilayers as a
function of temperature, T. Experiments performed at different times on
surfaces of different size: ($\circ$) 5x5 cm$^2$ ($\bullet$) 5x8cm$^2$.
Both experiments were performed on D16.(b) Distance, D$_w$, between two
Di-C$_{18}$-PC bilayers. Experiments performed at different times over two
years on different instruments: ($\square$) sample A, measured on D17;
($\bigtriangledown$) sample A, second time raised in temperature after
cooling from 58$^\circ$C; ($\circ$) sample B, measured on D16;
($\bigtriangleup$) sample C measured on D16; ($\Diamond$) sample D measured
on D16  \protect\cite{fragneto}.}
\end{figure}

\begin{figure}
\label{reversi}
\caption{(a) Distance, D$_w$, between two Di-C$_{17}$-PC bilayers as a
function of the temperature, T: ($\Box$) first time up in T; ($\bullet$)
first time down in T; ($\Diamond$) second time up in T. (b) Distance, Dw,
between two Di-C$_{18}$-PC bilayers for 2 samples. Sample 1: ($\square$)
first time up (1); ($\blacksquare$) first time down (2);  ($\Diamond$) second time up (3);
Sample 2: ($\circ$) first time up (1$^\prime$); ($\bullet$) first time down (2$^\prime$). All
experiments were performed on D17.}
\end{figure}

\begin{figure}
\label{compasigma}
\caption{Comparison of values of the r.m.s. roughness for a Di-C$_{16}$-PC
sample heated from 25 to 82.5$^{\circ}$C as a function of temperature, T
($\square$) as determined experimentally by model fitting the data and
($\bullet$) as extracted with the self-consistent theory
\protect\cite{mecke}.}
\end{figure}

\begin{figure}
\label{kappa}
\caption{Values of the bending modulus $\kappa$ normalised by $k_B T$ as a
function of the deviation of temperature from the melting temperature,
T-T$_m$ for the samples Di-C$_n$-PC with ($\bullet$) n=16; ($\circ$) n=17;
($\blacksquare$) n=18; ($\square$) n=19; ($\lozenge$) n=20. Values of
$\kappa$ are normalised to the arbitrary value 10$k_B T$ in the fluid
phase. This method forces the values for all different lipids to overlap
artificially in the fluid phase. The dashed line represents the
pretransition temperature T$_p$.}
\end{figure}

\newpage

\onecolumn

\begin{figure}[h]
\vspace{-1truecm}
\centerline{\epsfxsize=15truecm \epsfbox{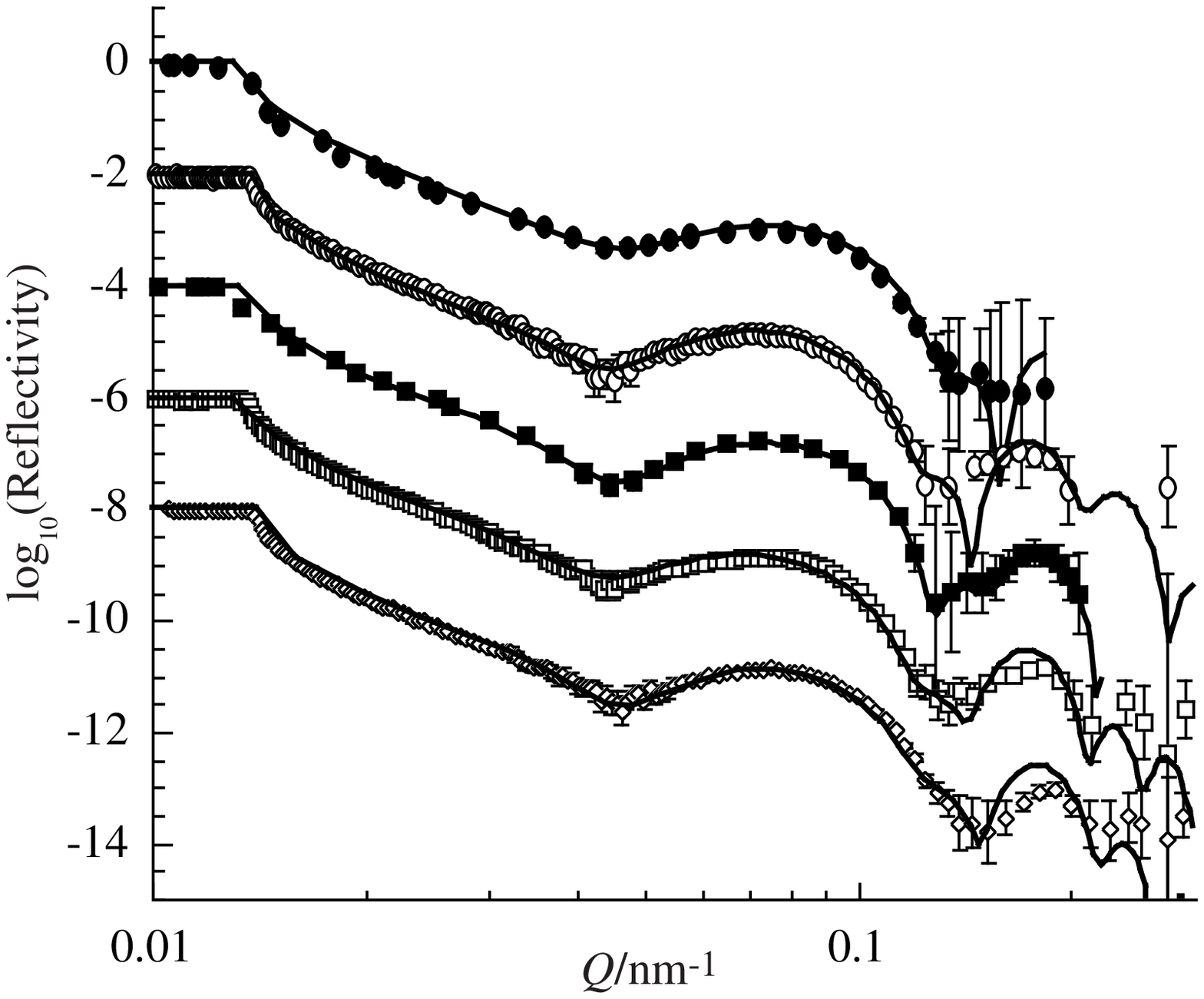}}
\vspace{-1truecm}
\end{figure}

\newpage

\begin{figure}[h]
\vspace{-1truecm}
\centerline{\epsfxsize=15truecm \epsfbox{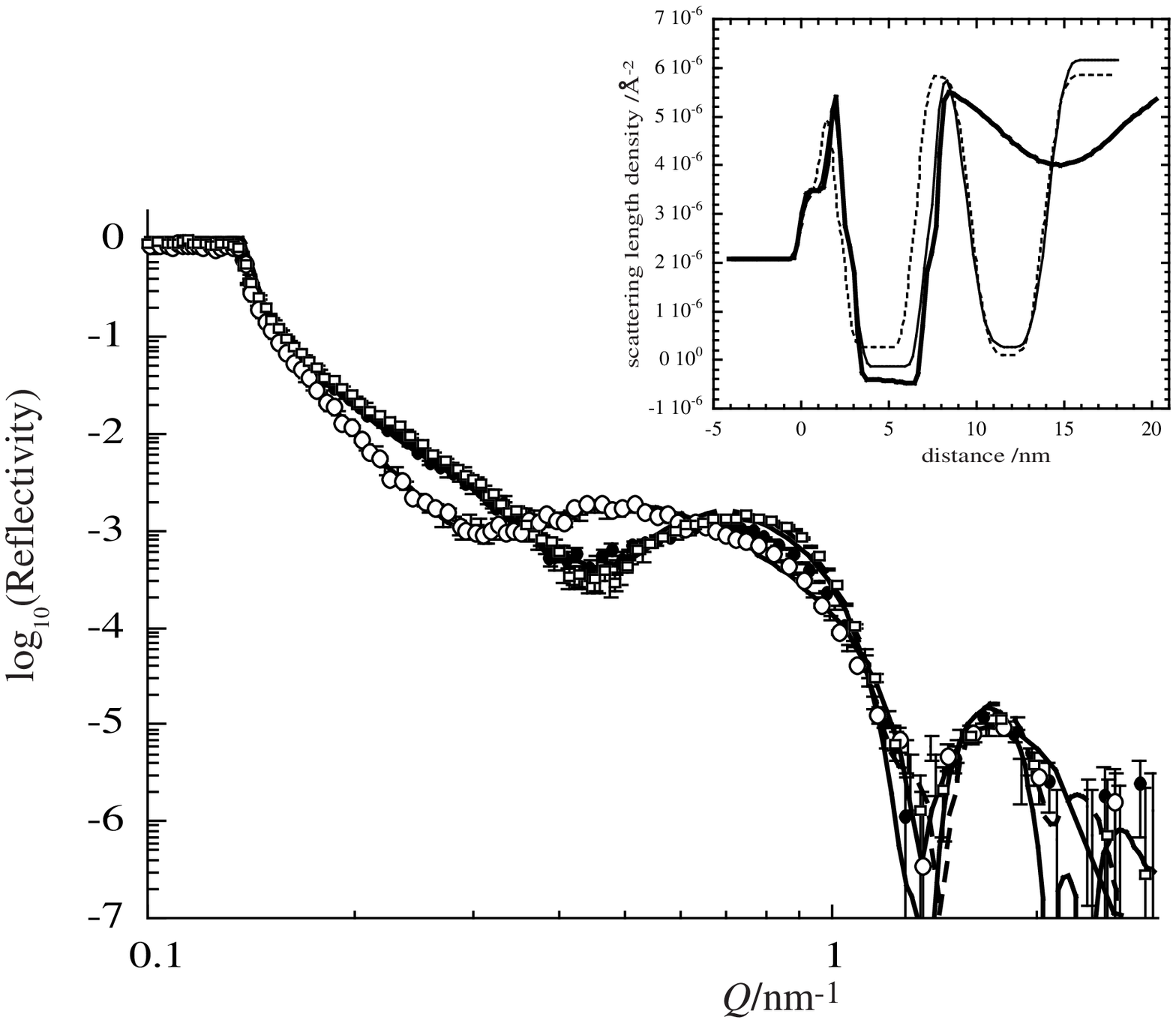}}
\vspace{-1truecm}
\end{figure}

\newpage

\begin{figure}[h]
\vspace{-1truecm}
\centerline{\epsfxsize=15truecm \epsfbox{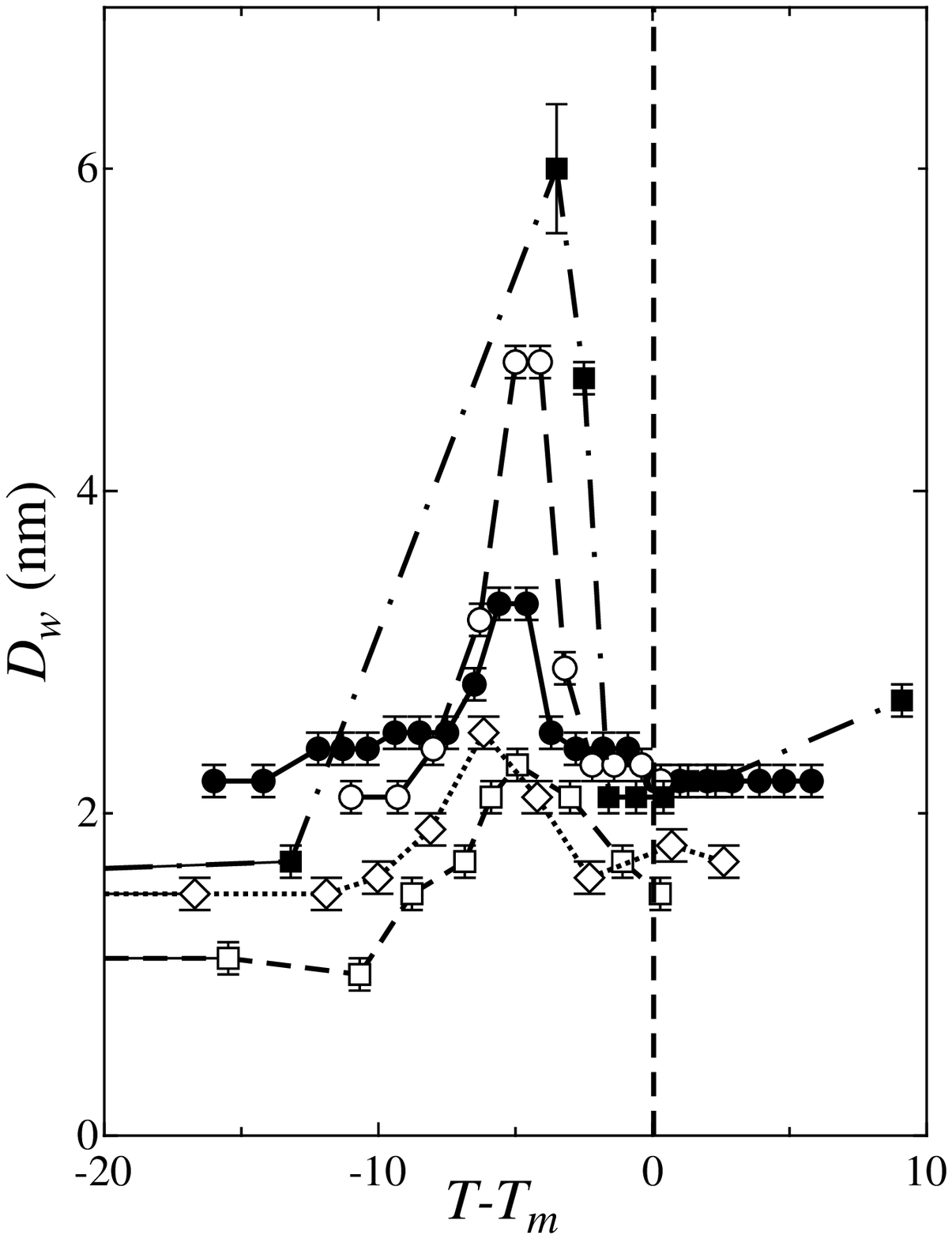} }
\vspace{-1truecm}
\end{figure}

\newpage

\begin{figure}[h]
\vspace{-1truecm}
\centerline{\epsfxsize=15truecm \epsfbox{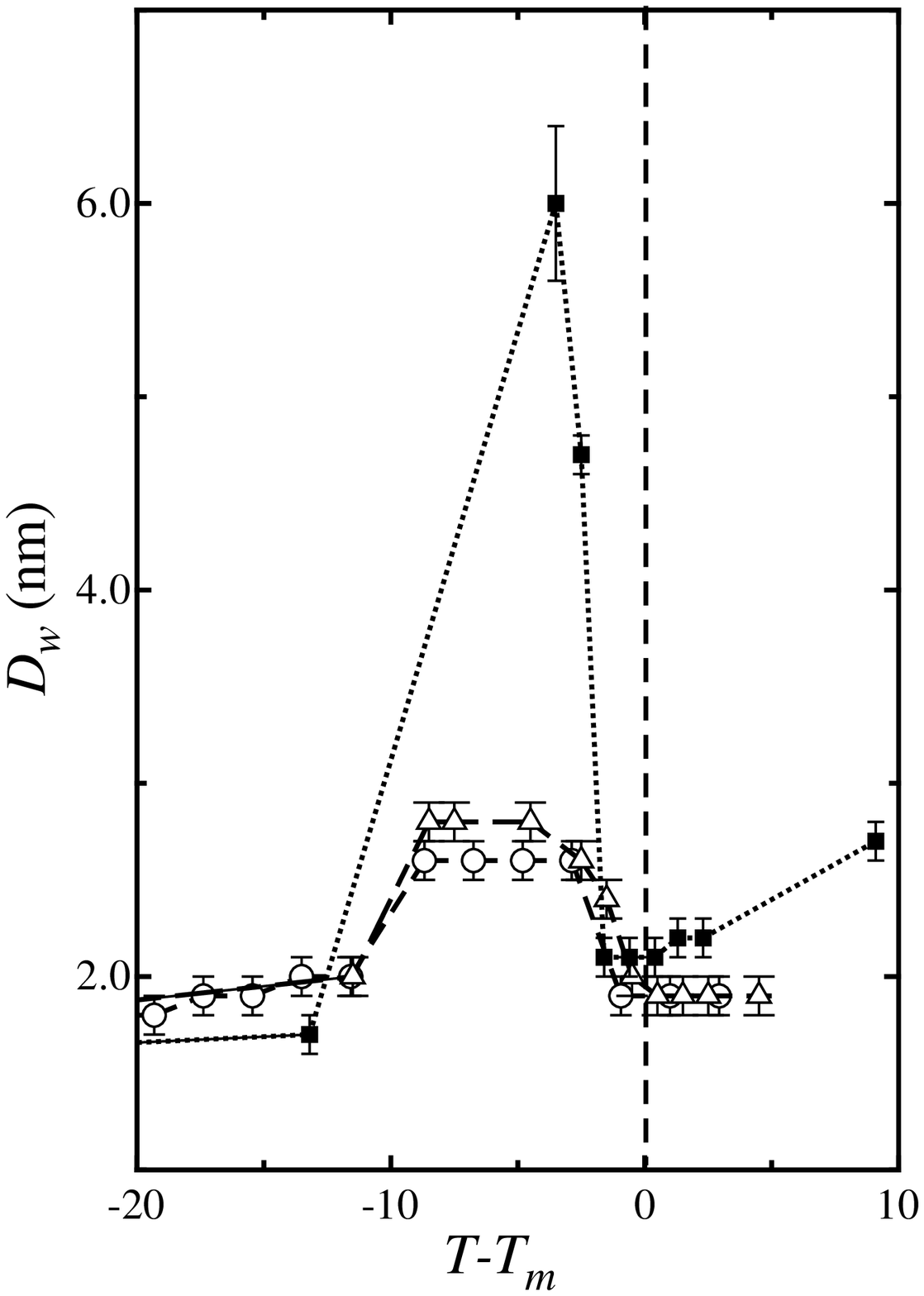} }
\vspace{-1truecm}
\end{figure}

\newpage

\begin{figure}[h]
\vspace{-1truecm}
\centerline{ \epsfxsize=15truecm \epsfbox{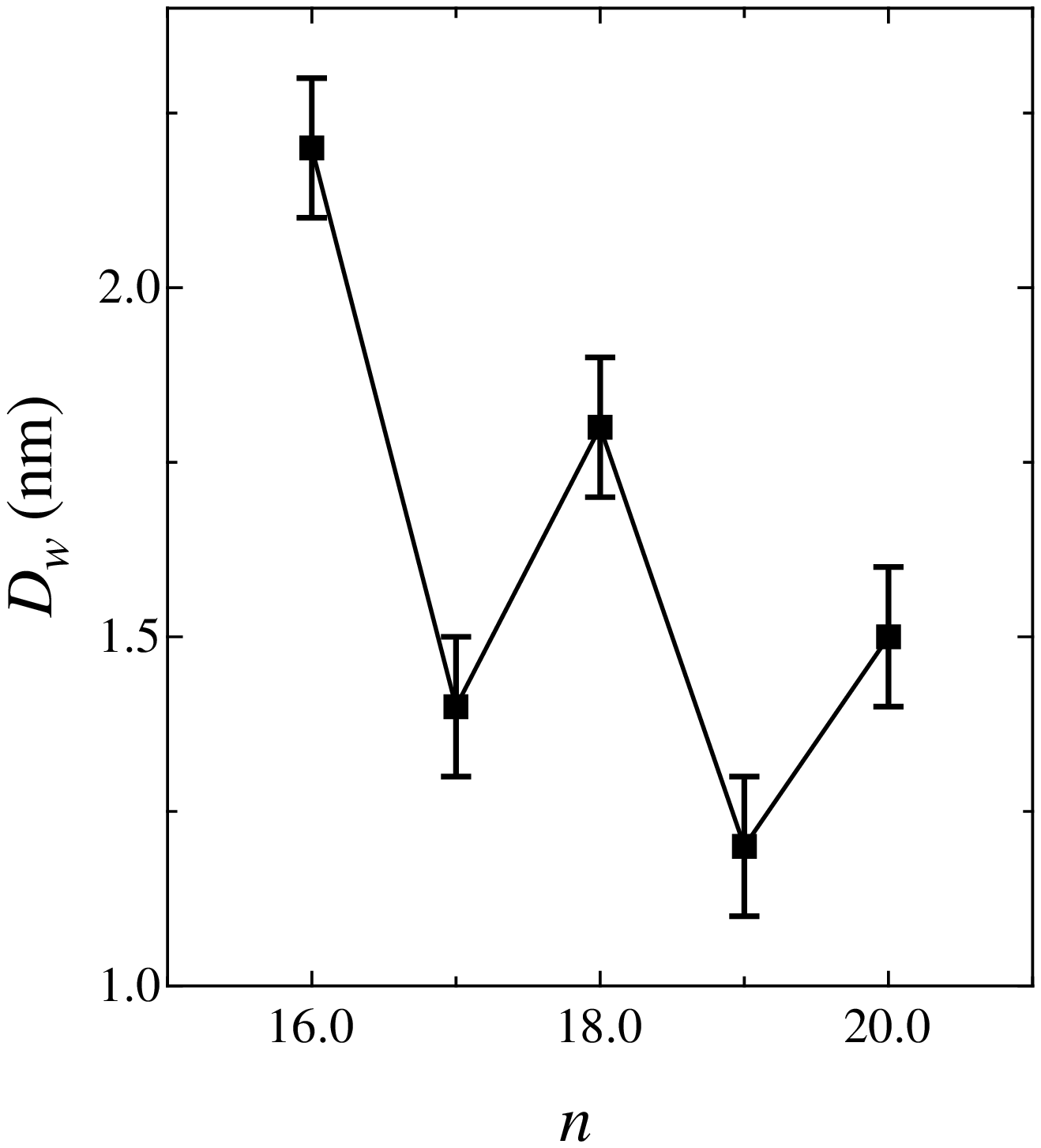} }
\vspace{-1truecm}
\end{figure}

\newpage

\begin{figure}[h]
\vspace{-1truecm}
\centerline{ \epsfxsize=15truecm \epsfbox{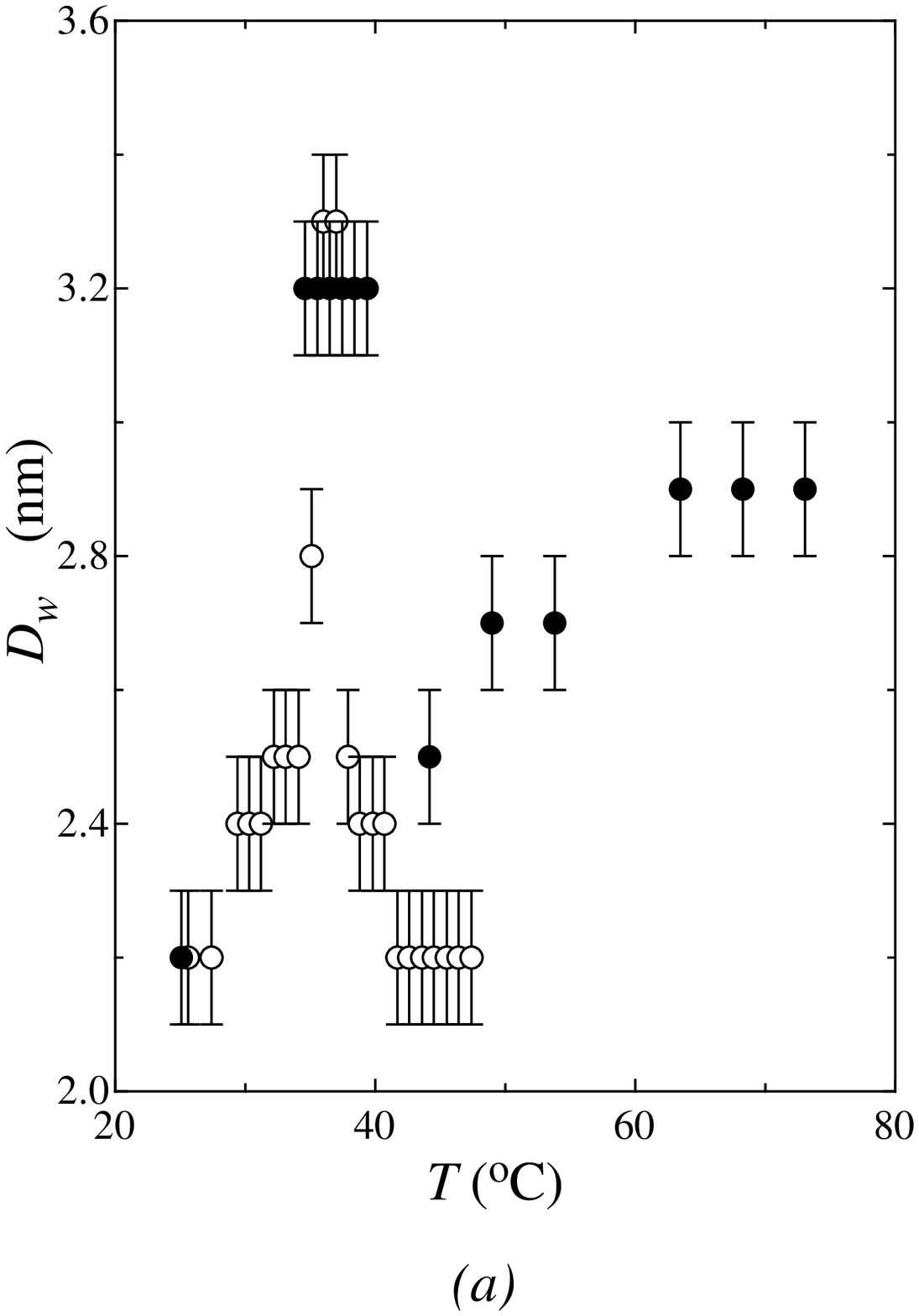} }
\vspace{-1truecm}
\end{figure}

\newpage

\begin{figure}[h]
\vspace{-1truecm}
\centerline{ \epsfxsize=15truecm \epsfbox{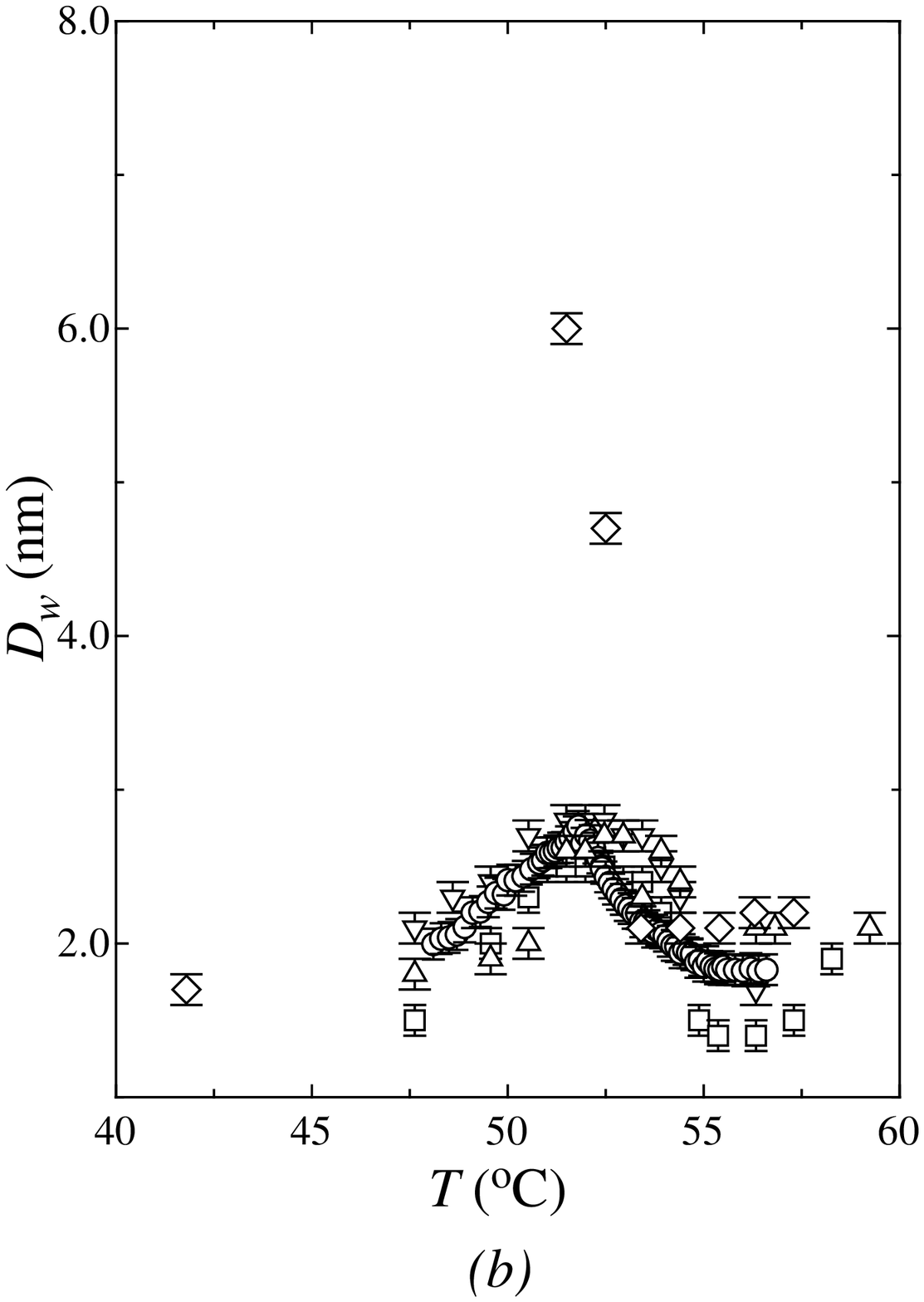} }
\vspace{-1truecm}
\end{figure}

\newpage

\begin{figure}[h]
\vspace{-1truecm}
\centerline{ \epsfxsize=15truecm \epsfbox{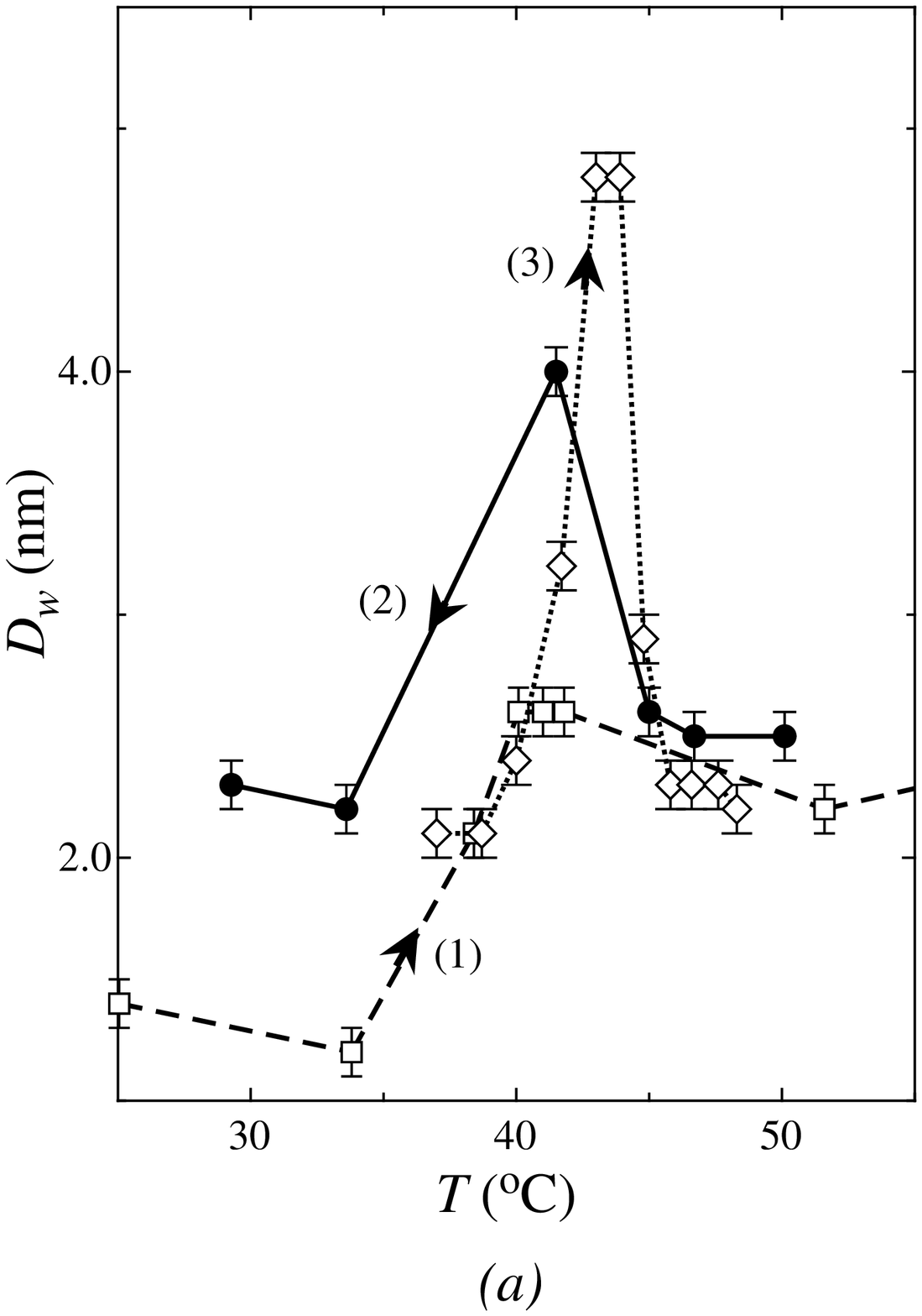} }
\vspace{-1truecm}
\end{figure}

\newpage

\begin{figure}[h]
\vspace{-1truecm}
\centerline{ \epsfxsize=15truecm \epsfbox{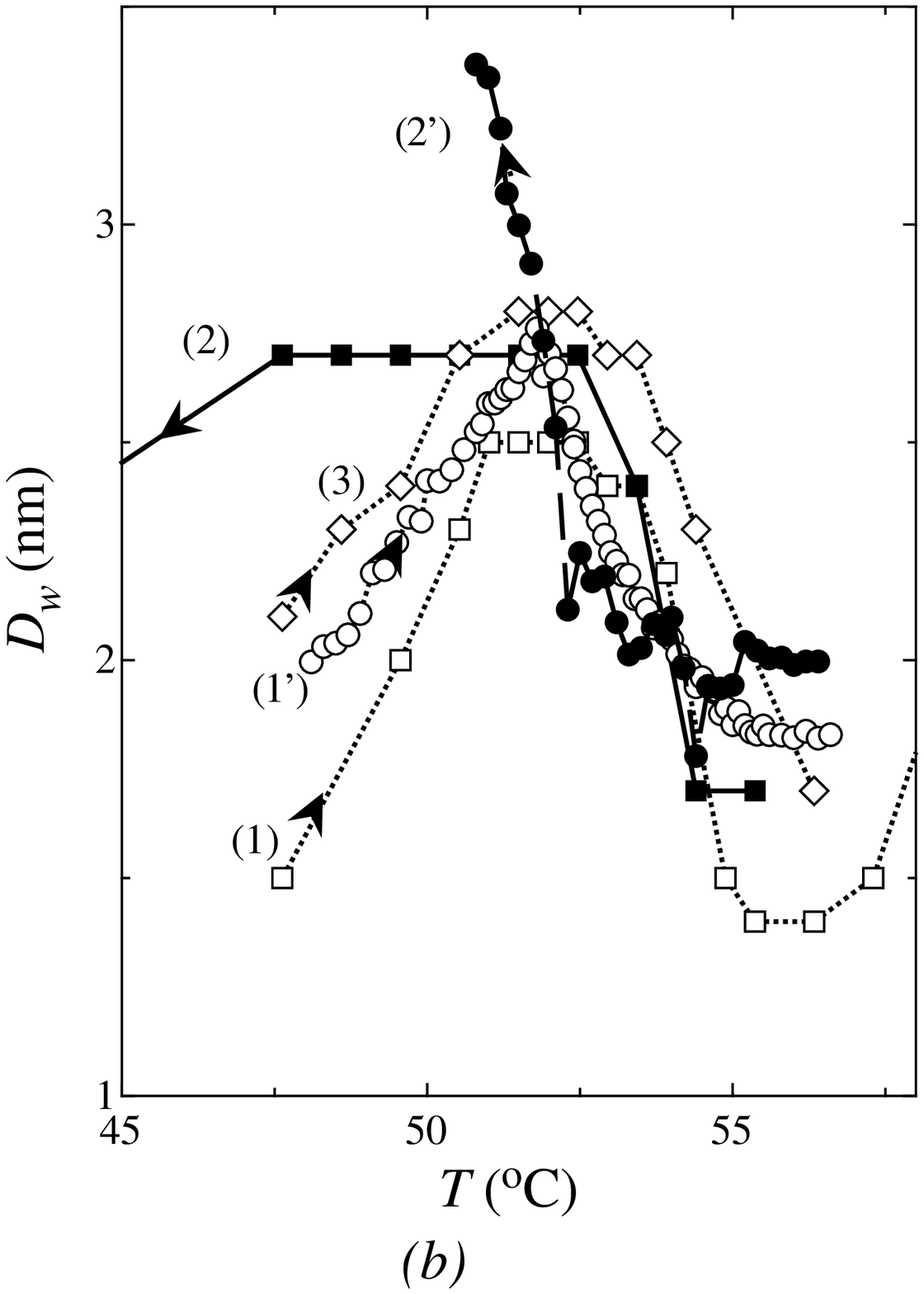} }
\vspace{-1truecm}
\end{figure}

\newpage

\begin{figure}[h]
\vspace{-1truecm}
\centerline{ \epsfxsize=15truecm \epsfbox{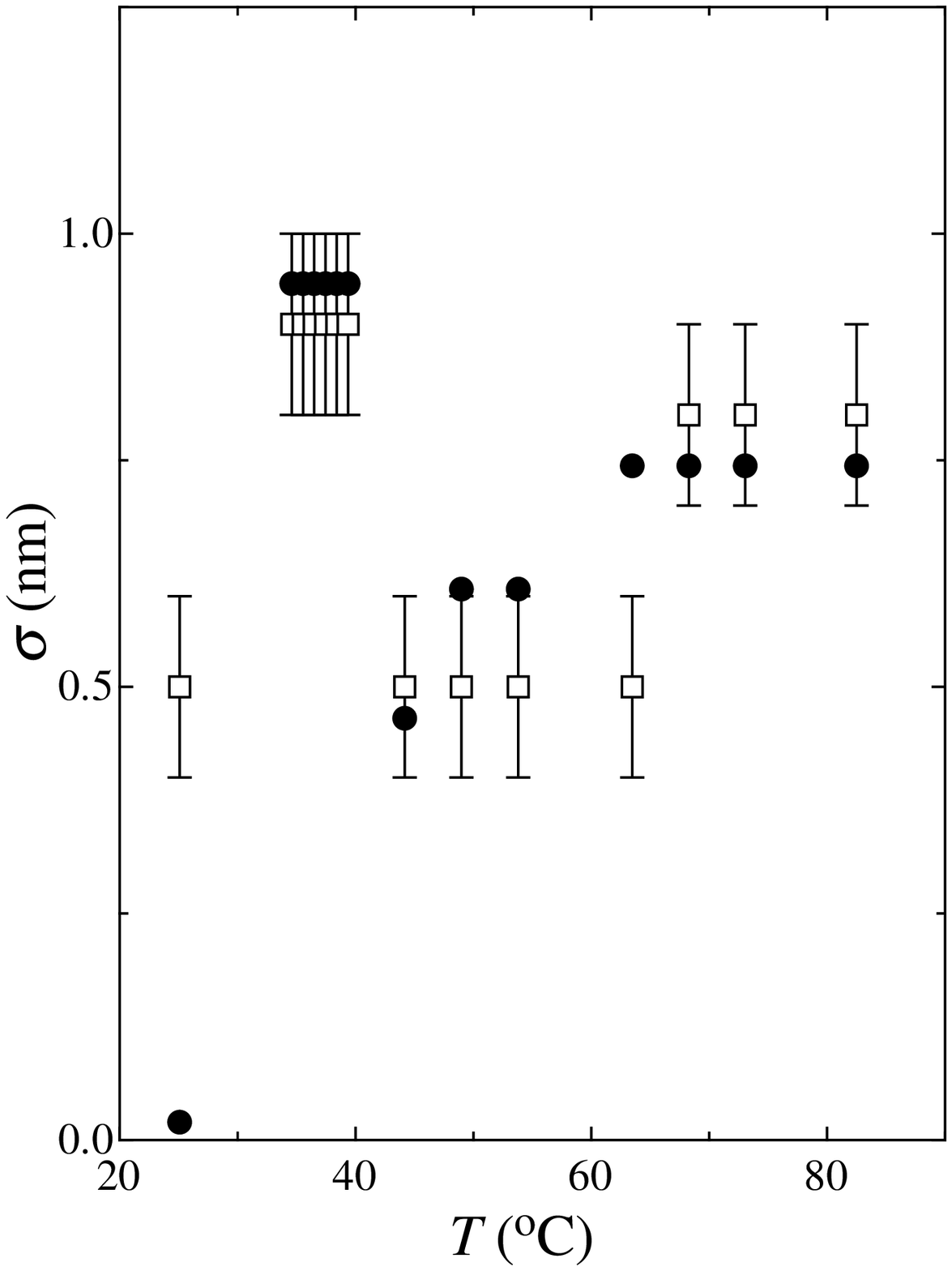} }
\label{figtenstrapped}
\vspace{-1truecm}
\end{figure}

\newpage

\begin{figure}[h]
\vspace{-1truecm}
\centerline{ \epsfxsize=13truecm \epsfbox{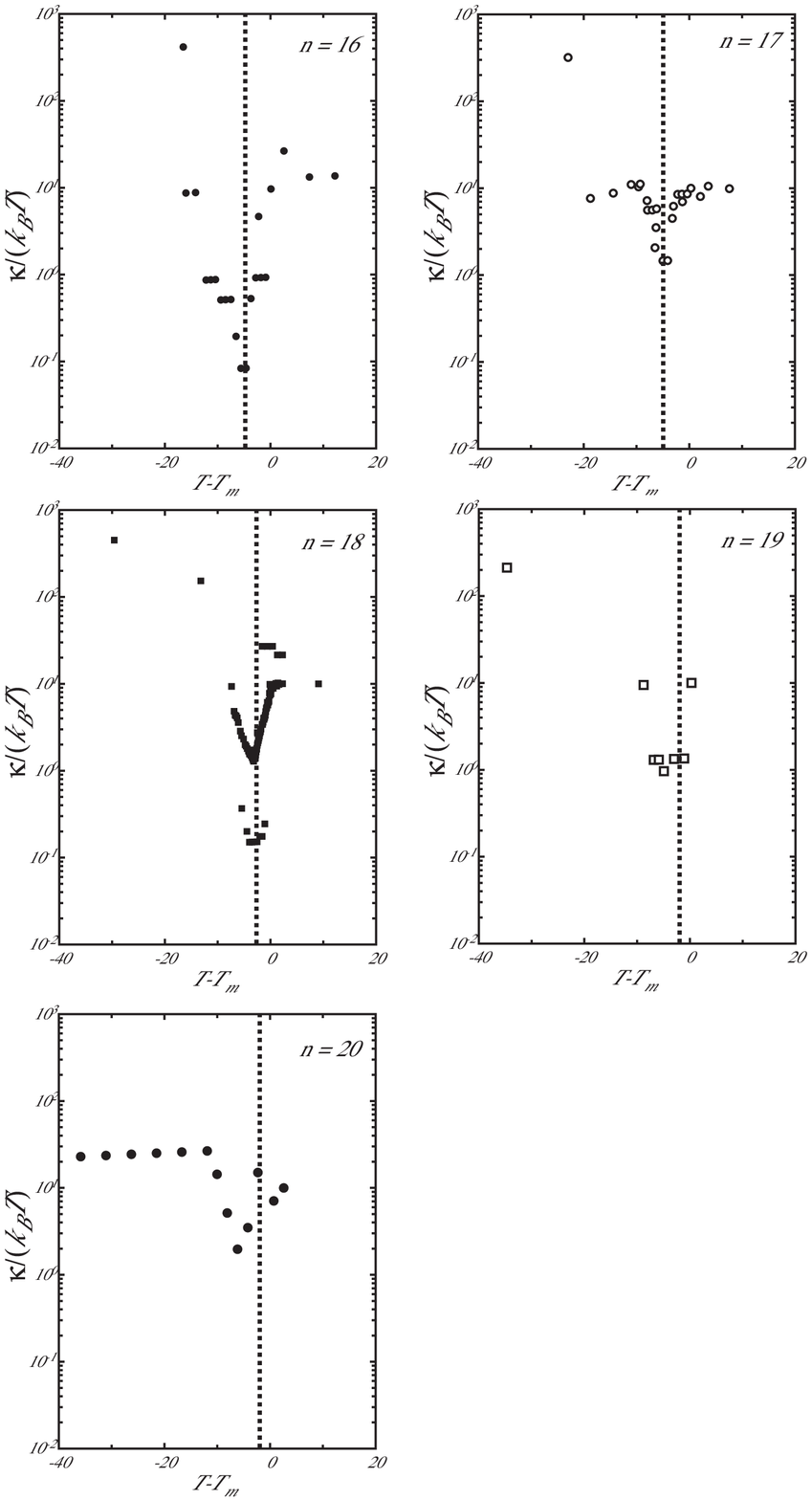} }
\vspace{-1truecm}
\end{figure}

\end{document}